\documentclass{article}

\PassOptionsToPackage{numbers, compress}{natbib}

\usepackage[creativeai, preprint]{neurips_2026}

\usepackage[utf8]{inputenc} 
\usepackage[T1]{fontenc}    
\usepackage{hyperref}       
\usepackage{url}            
\usepackage{booktabs}       
\usepackage{float}          
\usepackage{amsfonts}       
\usepackage{microtype}      
\usepackage{graphicx}       
\usepackage{xcolor}         

\title{Afterlife Delegation Protocol: Speculative Design of Self-Sovereign Agents that Outlive Their Principals}

\author{%
  Botao Amber Hu\\
  University of Oxford\\
  Oxford, UK \\
  \texttt{botao.hu@cs.ox.ac.uk}
  \And
  Iris Long \\
  Goldsmiths, University of London \\
  London, UK \\
  \texttt{irisxink@gmail.com}
}

\begin{document}

\maketitle

\begin{abstract}
\textit{Afterlife Delegation Protocol} is a speculative design project that asks what death becomes
when a will can act eternally. We design a speculative protocol through which a living person signs an
\textit{agentic will}: upon
a verified death, a self-sovereign AI agent spawns on blockchain --- an immutable, resistant,
decentralized, infrastructural substrate that could last forever --- endowed
with the funds and memories its
principal attached to it, and persists indefinitely to execute the will, overridable by no
custodian. Rather than argue about this future, we stage it: following the science fiction science
method, we translate the speculation into an \textit{experiential futures} intervention ---
a working web platform where real people design their own afterlife agents through an iterative,
interactive, AI-automated interview, re-login to revise, and rehearse their will in a sandbox. Their
drafted wills become qualitative data on a question rarely askable directly --- \textit{what should
outlive you?} --- and on how afterlife cosmologies across different cultural beliefs --- Buddhist,
Christian, Hindu, Muslim, and atheist --- begin to drift under the pressure of AI proliferation. We
design the protocol over a composition of existing Ethereum agent standards --- drafted as \texttt{ERC-10001}, in
the normative format of an Ethereum Improvement Proposal --- and describe its three-stage lifecycle
(designing the afterlife, proof of death, agent enactment), the research method, and preliminary
observations from an ongoing collection. The work surfaces a poetic
delegation moment between human mortality and machine eternality, mediated by long-lived
infrastructures that span generations.
\end{abstract}

\section{Introduction}

\begin{quote}
\textit{People die twice. First, when they die. Then, when they are forgotten.} --- Rokusuke Ei,
1933--2016 \citep[p.~62]{ei1995nidome}
\end{quote}

\noindent A will is the oldest instrument for exercising agency in a future where one cannot act.
In every existing form of will execution --- a testament interpreted by lawyers and
relatives \citep{friedman2009deadhands}, a trust administered by mortal trustees
\citep{simes1955deadhand}, an artist's estate kept by a foundation \citep{lydiate2007posthumous} ---
the will is a static declaration that runs on institutional permission and ceases to act at the
boundary of institutional attention. This project speculates a stronger instrument. In the future we design, a person signs an \textit{agentic will} while alive; after
their death is verifiably witnessed, the will spawns a dedicated, self-sovereign AI agent
\citep{hu2025trustless} --- an executor that is not a person but a persistent autonomous process. The agent can be endowed with
funds (a treasury from which it pays for its own compute, storage, and commissioned work) and with
memories (archives, writings, likenesses, and the recorded interviews through which the will was
elicited). It then acts in the world indefinitely: maintaining, responding, transacting, refusing,
on behalf of someone who can never again be consulted.

\begin{quote}
  \textbf{As a mortal being, what do you believe should outlive you?} Your social identity? Your assets? Your memories? Your trace of living? Your reputation? Your unfinished wishes? What if a self-sovereign AI agent could carry out your afterlife on an almost ethereal infrastructure?
\end{quote}

This is a speculative design project \citep{dunne2013speculative} --- a provotype
\citep{mogensen1992provotyping,boer2012provotypes}, not a product prototype. But it departs from traditional
scenario-writing: it is a high-fidelity simulation. Technologies of this kind sit inside the
Collingridge dilemma \citep{collingridge1980social} --- their impacts cannot be known until they are
developed and deployed, yet by then they are entrenched and hard to steer. To gain knowledge of
impact before entrenchment, we deploy Rahwan et al.'s \textit{science fiction science} method
\citep{rahwan2025scifi}: rather than predicting whether such agents will exist, we build a
functioning simulacrum of the future and study how people behave inside it, as the Moral Machine
did for autonomous vehicles \citep{awad2018moral}. We then translate the simulation into an
\textit{experiential futures} intervention \citep{candy2010futures,candy2017experiential}: a real,
public web platform where visitors do not read
about afterlife agents but encounter one --- they sit for an interactive, iterative long interview
with an automated AI interviewer \citep{anthropic2025interviewer}, draft their own posthumous
delegation, revise it, and watch a sandboxed rehearsal of their agent acting after their
hypothetical death.

What the provotype measures is cultural drift. Mourning, reincarnation, judgment, inheritance,
ancestor veneration, and the idea of a second death are cultural technologies for bringing the
agency of the dead to a close \citep{hertz1960death,aries1981hour}. AI proliferation puts that
closure under pressure, and different afterlife cosmologies
will bend differently: each tradition holds a distinct cosmological model of what follows death
\citep{bowker1991meanings,obayashi1992death,walls2008eschatology,moreman2017beyond}. So what would
a Buddhist, a Christian, a Hindu, a Muslim, and an atheist each believe should outlive their
original life on-chain, eternally, with AI agents?
The project collects these answers as they begin to form, before the technology hardens them.

The paper contributes: (1) the speculative design of the protocol itself; (2) its staging as an
experiential future for real participants; and (3) preliminary observations and the cross-cultural
probes of an ongoing collection.

\section{Background}

\subsection{AI, afterlife, and immortality}

Terror management theory holds that people buffer the knowledge of their own death by embedding
themselves in structures that outlast them: lineage, religion, nation, institution, work
\citep{greenberg2012terror}. Every such structure carries a cosmology of time --- an implicit account
of how long things last, what deserves to persist, and who is owed continuity
\citep{munn1992time}. Continuing bonds research shows that the living already sustain
relationships with their dead without any technology at all \citep{klass1996continuing}; the
digital afterlife industry now supplies technology in quantity \citep{ohman2018afterlife}. A
growing literature maps it:
responsible design for griefbots and deadbots \citep{hollanek2024griefbots}, \textit{generative
ghosts} as agentic AI representations of the dead \citep{morris2025generative}, TeleAbsence as a
poetics of mediated absence \citep{ishii2025teleabsence}, design research into AI carriers for
afterlife selves \citep{lai2026horcrux}, and studies of how people perceive an ``AI afterlife''
as digital legacy \citep{lei2025aiafterlife}. In Morris and Brubaker's design space, the artifact we
speculate is a first-party, pre-mortem, evolving ghost --- but it adds a dimension their framework
does not cover: \textit{infrastructural persistence} (Section~\ref{sec:designspace}). Existing
ghosts are products, hosted at a company's pleasure; they die when a server bill lapses. The
question of what happens when the representation of the dead cannot be turned off has, until
recently, been unaskable.

\subsection{AI, blockchain, and immutability}

Decentralized infrastructure makes it askable. A public blockchain is a machine for immutability:
state that no single party can roll back, contracts without custodians, funds that disburse by rule
rather than by permission. This is the Long Now \citep{brand1999longnow} arriving as an executable
substrate rather than a clock, with all the maintenance politics that infrastructural longevity
actually entails \citep{ribes2009longnow,brand2025maintenance}. Two further ingredients make the substrate habitable by
agents. Decentralized physical infrastructure networks (DePIN)
\citep{lin2025depin,ballandies2023depin} now provide \textit{permissionless compute}: computation
that can be bought by anyone --- including a contract --- without an account, a company, or a
jurisdiction consenting. And confidential computation in trusted execution environments (TEEs)
\citep{li2023teesurvey,costan2016sgx}, as proposed for TEE--EVM co-processing in ERC-733
\citep{erc733tee}, lets that computation run privately while proving on-chain \textit{which} code
ran. Neither is hypothetical: crypto-native services such as
Venice.ai already sell open-weight foundation-model inference for cryptocurrency, permissionlessly
and without accounts, and TEE cloud networks such as Phala make such compute attestable
on-chain\footnote{\url{https://venice.ai}; \url{https://phala.network}.} --- an agent can, today, buy
its own inference. Deployed together, these already support LLM-based agents
that operate ``in the wild'' --- holding their own wallets and social accounts, paying for their own
inference, with no human able to reach in and stop them: the Spore.fun experiment documented
sovereign agents evolving on TEE-secured blockchains as a live, open-environment system rather than
a thought experiment \citep{hu2025spore}. Such an agent inherits its non-overrideability from the
infrastructure itself rather than from any custodian's forbearance \citep{hu2026sovereign}. The
combination is what makes our speculation load-bearing: an agent that is at once
\textit{autonomous} (AI), \textit{unstoppable} (permissionless compute)
\citep{hu2024unstoppable}, and \textit{unforgetting}
(immutability) is the first vessel of symbolic immortality \citep{lifton1979broken} that can also
act. The cultural
technologies of death \citep{hertz1960death,aries1981hour} have never had to metabolize that
before.

\section{Speculative design: the Afterlife Delegation Protocol}

We outline the protocol as \texttt{ERC-10001}, published at \url{https://erc10001.org} in the exact
normative format of an Ethereum \citep{buterin2014ethereum} Improvement Proposal. It is built on
four existing Ethereum agent standards, each supplying one organ of the afterlife agent: ERC-733 \citep{erc733tee} the \textit{sealed
skin} --- an attested confidential boundary (TEE--EVM co-processing) inside which the agent's
\emph{body} (the compute harness) and \emph{mind} (the foundation model) run inviolate, invadable
and alterable by no
outsider, not even the machine's operator; ERC-8004 \citep{eip8004} the \textit{identity} --- the ID
through which the sealed agent faces outward, a persistent, discoverable registration with
reputation and validation registries that lets it exchange money and information with the external
world; ERC-8350 \citep{erc8350} the \textit{memory} --- a memory state registry anchoring every
evolution of the agent's memory in a strictly linear hash chain, so that memories of the dead
cannot be silently deleted, rolled back, or forked into contradictory ghosts, and so that death
itself appears in the ledger as a change of author within one unbroken chain; and ERC-8183
\citep{eip8183} the \textit{metabolism} --- job-based escrow through which the agent buys storage,
commissions work, and hires other agents, funds leaving the treasury only against attested
delivery. Above these, the protocol adds what none of them contemplate: a
principal who dies. It enforces a constitutional separation between the signed \textit{core will},
immutable at death, the agent's accumulating operational memory, and the model substrate
interpreting both --- the agent may become more capable; it may not become someone else. The
lifecycle runs in three stages; the full specification is given in
Appendix~\ref{app:spec}.

\paragraph{Stage 1: Designing the Afterlife.} The first stage is alive today as a working platform at
\url{https://afterlife-protocol.org}. A visitor does not fill in a form; they sit for a long,
interactive, AI-automated interview with the \textit{Afterlife Interviewer}
(Appendix~\ref{app:interviewer}), which converts diffuse intuitions about death into specific,
contestable commitments: what the agent may say and to whom, in which pronoun (\textit{as} the
deceased or only \textit{about} them), with what funds, under which prohibitions, for how long, and
what may terminate it. Two properties make the stage a design environment rather than a survey. The
will is \textit{revisable}: participants re-login across weeks and change their minds, every
revision a signed, versioned commit --- a will here is not a final declaration but a living
negotiation with one's own future absence. And the will is \textit{sandbox-testable}: a simulation
lets participants converse with their own agent after their hypothetical death, watch it handle
scripted situations (a grieving relative, an impersonation attempt, an unforeseen request), and
return to revise the clauses that failed them. The
current deployment is a research prototype: wills are drafted and simulated, never activated, and
no agent of any actual deceased person is built.

\paragraph{Stage 2: Proof of Death.} Everything downstream hangs on one predicate the chain can
never compute: that the principal has died. A \textit{death oracle} must witness the death --- and
death is the unique event that cannot be self-attested; liveness is self-evidencing, death is only
ever hearsay. The protocol therefore treats attestation as its most sensitive interface and designs
it as a weighted, contestable proceeding rather than a switch: any party may file a bonded death
claim; evidence accumulates weight from will-designated witnesses (institutional registries, human
witness quorums, the lengthening silence of a dead-man's switch); a challenge window scales
inversely with the weight of evidence; and one invariant is supreme --- any authenticated sign of
life from the principal vetoes everything. The protocol may wrongly delay a ghost; it must never
overrule a living voice. Crucially, the definition of death is itself testamentary content:
cardiac, brain, legal, social, and ``key'' death are different events, ranked differently by
different traditions, so the will configures which deaths count, witnessed by whom, on what
horizon.

\paragraph{Stage 3: Agent Enactment.} Upon a finalized attestation, the will becomes immutable, the
treasury passes to the agent, and the agent begins acting on-chain in the real world through a
persistent loop: read the will, interpret its obligations, observe the world, plan, act, record
memory, sustain itself. It maintains the will and acts on the principal's unfinished goals ---
tending archives and social presences, responding to designated people, protecting the deceased's
likeness against misuse, funding memorial rituals, continuing incomplete projects, and hiring other
agents for what it cannot do alone. It may upgrade to newer models while the original will binds it
as a constitution, and it may be terminated only through the governance mechanism the principal
consented to in advance. Because its endowment spends at most its yield, decline is graceful rather
than sudden, and anyone --- a grandchild, a stranger, an institution --- may add funds at any time.
The living can feed the ancestor.

\section{Research method: staging the future to collect it}

We stage Stage~1 as a science fiction science instrument \citep{rahwan2025scifi} to engage online
participants, collecting their interactions with afterlife agents. The framing matters: the
experience presents itself as delegating one's will to an afterlife-agent \textit{service}, and the
interview behaves as consulting and sales interviews do, eliciting requirements from a client who
does not yet know what they want --- most people cannot fully name their will on first reflection,
so articulation is necessarily iterative, explorative, and qualitative. Asked directly, \textit{what should outlive you?} yields platitudes; asked as
\textit{what may your agent do, say, spend, and refuse in year forty?} --- and asked again, across
sessions, against the sandbox's rehearsals --- it yields commitments, each clause a datum about what
its author believes deserves continuity, who is owed it, and how long ``forever'' is felt to be,
complemented by qualitative surveys on what people believe should persist beyond a life. The
interviews are conducted by the Afterlife Interviewer, an adaptive semi-structured interviewing
agent built on SparkMe \citep{anugraha2026sparkme}, which plans questions, takes notes, and
pursues emergent subtopics --- scaling qualitative elicitation to an online population without a
human interviewer (Appendix~\ref{app:interviewer}).

The ongoing collection focuses on cross-cultural difference across beliefs and faiths --- Buddhist,
Christian, Hindu, Muslim, and atheist participants are recruited purposively --- because the
project's object is precisely the cross-cultural drift: how differently afterlife cosmologies
\citep{moreman2017beyond} absorb the offer of a machine executor. Because the material concerns grief, private memory, and
named third parties, transcripts are never published; analysis draws on Clio
\citep{tamkin2024clio}, Anthropic's privacy-preserving pipeline for insights from real-world AI
conversations, to surface population-level themes while raw responses remain sealed
(Appendix~\ref{app:interviewer}). Participation is voluntary and consent-based; participants may
withdraw and delete their wills.

\section{Preliminary results}
\label{sec:results}

The collection is early and ongoing; we report 4 early qualitative observations.

\paragraph{Reputation maintenance: the curated self, defended forever.} The protocol invites
estates and legacies; what participants first ask for is often
something small and exact. One participant, after a long interview, settled on a single standing
instruction: the agent should maintain their social
media photos forever --- including automatically correcting and ``photoshopping'' new and old
images so the feed never falls out of date: \textit{``keep my photos the way I keep them now. If
a picture of me is bad, fix it. I don't want to look worse dead than I did
alive.''} The clause is easy to smile at and hard to dismiss: it is continuing impression management, a
curated self defended after death with the care it was defended in life --- suggesting that what
people first reach for, when offered infrastructural immortality, is not wealth or wisdom but the
integrity of a self-image. Against the literature's grand ghosts, the vernacular afterlife may be
maintenance work.

\paragraph{Wiring money to the ancestor (Chinese ancestor veneration).} Among participants from
Chinese ancestor-veneration backgrounds, persistence is welcomed as the extension of a practice
already millennia old \citep{ahern1973cult,wolf1974gods}. One participant asked the agent to keep the family's calendar of rites: \textit{``at Qingming it
should greet my grandchildren and bless them for the year. We have always fed our ancestors ---
but joss paper burns, and the money never arrives. If my grandchildren wire crypto to it, that
isn't a deposit --- that's an offering, and for the first time the ancestor can actually
spend it: on its compute, its maintenance, next year's blessing.''} The ritual economy becomes
literal: offerings once burned now arrive as funds the ancestor-agent spends to stay alive ---
purchasing its own compute and upkeep --- so its immortality depends on how faithfully the living
keep paying. The best-fed ancestors will live longest.

\paragraph{Dissolution versus the merit engine (Buddhist).} For Buddhist traditions built on
impermanence and non-attachment, an afterlife agent that cannot stop is a category error:
clinging to continuation is precisely what the doctrine trains one to release. At the same time,
in Buddhist cosmology merit (karma) accumulates across the whole cycle of rebirths, and making
merit on behalf of the dead is already established practice
\citep{gombrich1971merit,cuevas2007buddhistdead}; so wills appear instructing the agent to keep
generating good karma --- perpetual giving, sutra recitation, acts of care --- through every
incarnation of its author: \textit{``let it keep giving in my name, life after life --- whoever
I am born as next should arrive into the good it has already done.''} Each new life thus
inherits the accumulated merit of an on-chain agency it cannot remember signing. The clash is
the protocol's default itself: infrastructure assumes forever; doctrine demands letting go. The
one reconciliation participants found is to turn persistence into a merit engine --- and then
letting go is exactly what the machine defers, forever.

\paragraph{Speaking \textit{about}, never \textit{as} (Christian).} In Christian cosmology the
soul has gone elsewhere, so an agent that speaks \textit{as} the dead comes close to mediumship;
participants therefore draft strict pronoun clauses --- the agent may speak \textit{about} them,
never \textit{as} them: a memorial, not a séance
\citep{morris2025generative,hollanek2024griefbots}. Yet the same participants comfortably assign
the agent perpetual charitable duties, which fit the familiar category of stewardship. One
participant pushed this logic to its theological edge. Carrying a wrong they regretted, they
asked whether the agent could keep atoning for it: \textit{``I've done things I'm not proud of.
If it keeps paying people back and giving after I'm gone, does that still count as me? Can it
change what's waiting for me?''} Doctrine says no: judgment is fixed at death, and no work can be
added afterward. But purgatory and prayers for the dead leave a door ajar, and through it the
participant volunteers the agent as a perpetual penitent --- importing into a cosmology of final
verdicts a karmic logic in which good deeds accumulate and slowly change the outcome. The drift
lies in the question itself: no longer whether the dead can act --- the infrastructure settles
that --- but whether their acts can still be credited to them.

\section{Discussion}

\paragraph{Cross-cultural belief drift under the pressure of AI.} Each tradition maintains a
distinct cosmological model of what follows death --- rebirth, resurrection and judgment, the
soul's onward journey, ancestral presence, extinction
\citep{bowker1991meanings,obayashi1992death,walls2008eschatology,moreman2017beyond}. An agency
that lasts forever does not merely extend these cosmologies; it conflicts with them.
Reincarnation presupposes that the self vacates this world; judgment presupposes a life whose
account has closed; an agent that goes on holding, spending, and speaking keeps the account open,
forcing questions the traditions never had to price: why rebirth, if agency persists here? Why
judgment, if the ledger never closes? Who answers for acts a will commits after its author can
no longer consent, repent, or be held to account? Like the Moral Machine \citep{awad2018moral},
the platform is a distributed cosmological probe --- but its dilemmas are unstandardized, and
what it collects is not convergence toward a shared answer but divergence in what the question is
taken to be. What social physics becomes in the age of AI --- the quantitative study of
collectives at once human and artificial \citep{han2026socialphysics,rahwan2019machine}, whose
culture machines increasingly generate, transmit, and select \citep{brinkmann2023machine} ---
will need instruments for such hybrid negotiations; this work is an early one.

\paragraph{Afterlife agency as a permissionless design space.}
\label{sec:designspace}
Because the executor is a general-purpose autonomous agent rather than a document, the space of
what a will can delegate widens from asset disposition to nearly anything an agent can be
instructed to do: an identity to be remembered by; a griefbot of living memory that
comforts the bereaved \citep{hollanek2024griefbots,klass1996continuing}; a knowledge system that
goes on contributing; a fortune that keeps compounding; a project scheduled in centuries; a
likeness defended against misuse; a dynamic, evolving ghost that carries a persona and continues to
act in society \citep{morris2025generative}. Morris and Brubaker
\citep{morris2025generative} give an early taxonomy of such ghosts along seven dimensions ---
provenance (first- vs.\ third-party), deployment timeline (pre- vs.\ post-mortem),
anthropomorphism paradigm (speaking \textit{as} vs.\ \textit{about} the dead), multiplicity,
cutoff date (static vs.\ evolving), embodiment, and representee type --- but the design space is
vaster still. And it is open in a second, self-sovereign sense: anyone may design and deploy an
afterlife agent without any platform's, church's, or state's permission, and once deployed on the
substrate, no custodian can stop it \citep{hu2024unstoppable,hu2026sovereign} --- whether this
future is found desirable or abhorrent, it is not gated.

\paragraph{Self-sovereignty vs.\ entrusting others to execute your agency.} Every prior
instrument of posthumous agency is an act of trust in the living: a testament trusts lawyers and
relatives to probate it \citep{friedman2009deadhands}; a trust trusts mortal trustees, capped by
the rule against perpetuities because the common law feared the dead governing indefinitely
\citep{simes1955deadhand}; moral rights trust heirs who thin out within a generation
\citep{merryman1976refrigerator,lydiate2007posthumous}; the artist's testament trusts an executor
who may refuse --- Kafka trusted Max Brod to burn the manuscripts, and Brod's refusal is why we
have \textit{The Trial} \citep{balint2018kafka}. The agentic will abolishes that dependence:
where legal tech automates the drafting and probate of wills \citep{crawford2020blockchain}, this
design automates the executor itself --- once activated, no relative, court, or platform can
rewrite or intervene. For the first time, agency can be genuinely protected after death:
entrusted to no one, diffused into infrastructure. Whether that is a good thing is exactly what
the design leaves open. Trust was the safety valve: the trustee's judgment, the court's equity,
and Brod's disobedience adapted dead instructions to living circumstances --- a faithful agent
would have lit the fire, and an unrevisable will that begins to harm the living can no longer be
reformed. And what each testator experiences as protection, society accumulates as sovereignty of
the dead, inheriting undiluted the dead-hand problem the law spent centuries containing. Do we
want agency protected even after death --- and who is ``we'' once the dead are still acting? That
question is what the termination-governance interface is for.

\paragraph{Danger of proof of death.} The deepest technical danger is that
\textit{proof of death is at least as hard as proof of humanity}. A decade of proof-of-personhood
work --- from pseudonym-party schemes \citep{borge2017pop} through live registries such as Proof of
Humanity \citep{kleros2021poh} to personhood credentials in the age of generative AI
\citep{adler2024personhood} --- asks how a digital system can verify that a unique living human
stands behind a key; our protocol needs the negation --- that no living human stands behind
it any longer --- and the negation is strictly harder, because liveness can demonstrate itself with
one signature while death can only ever be testified to by others. Prior designs for death
on-chain confirm the difficulty by avoiding it: blockchain will systems fall back on designated
human certifiers \citep{chen2021will}, dead-man's-switch protocols such as Sarcophagus reduce
death to the silence of missed check-ins \citep{sarcophagus2021}, and analyses of crypto
inheritance leave the death event itself as an open question \citep{prost2022inheritance}. Every
attestation path we compose (registries, witnesses, the lengthening silence of a dead-man's
switch) inherits the same trust assumptions, sybil risks, and biometric decay that personhood
systems face, plus failure modes of its own: a treasury payable on death is structurally a bounty on a life, coercion of
check-ins can be flagged but not prevented, and the vanished-but-alive are eventually mis-declared
by construction.

\paragraph{The transition moment between human mortality and machine eternality.} What the work
finally aims to surface is a poetic delegation moment: a mortal human, dictating terms to an
eternal machine, mediated by the quiet, embedded infrastructures through which civilizations may
persist across generations \citep{star1999ethnography,ribes2009longnow}. Blockchain immutability is a
maintained accomplishment \citep{graham2007outoforder,jackson2014rethinking,brand2025maintenance}
of validators, client developers, and node operators, so an agent's ``forever'' is willed anew by
each generation of hands; yet because that maintenance is decentralized, the substrate is
unstoppable in practice while contingent in principle \citep{hu2024unstoppable} --- no one in
particular keeps it alive, and no one in particular can turn it off. Social order quietly
presupposes finitude: institutions from inheritance to forgiveness are calibrated to actors who
die, and machines are the first social participants that need not \citep{liu2026timewithoutdeath}.
The agentic will places that asymmetry inside a single, signed relationship: whether the agent
keeps faith with its author is the question every heir has faced --- but for the first time, the
executor does not die either.

\section{Conclusion}

\textit{Afterlife Delegation Protocol} practices speculative design in the medium where
infrastructural futures are actually decided: the protocol specification. Drafting a plausible
standard, staging it as an experiential future, and collecting how differently people's
cosmologies answer it treats the speculative artifact not as a provocation to be exhibited but as
an instrument to be administered. The ghosts are coming either way; what remains designable is the
will that binds them, and what remains collectable is what the living actually want
to outlive them.

{\small
\bibliographystyle{abbrvnat}
\bibliography{reference}
}

\appendix

\section{Specification sketch: \texttt{ERC-10001} Afterlife Delegation Protocol}
\label{app:spec}

\textit{Supplementary material. The paper body ends with the references above; this appendix is
provided for reviewers who wish to inspect the technical proposal. The full normative draft, with
RFC-2119 language and the complete Solidity interface, is published at \url{https://erc10001.org}.}

\begin{figure}[H]
\centering
\fbox{\includegraphics[width=0.92\linewidth]{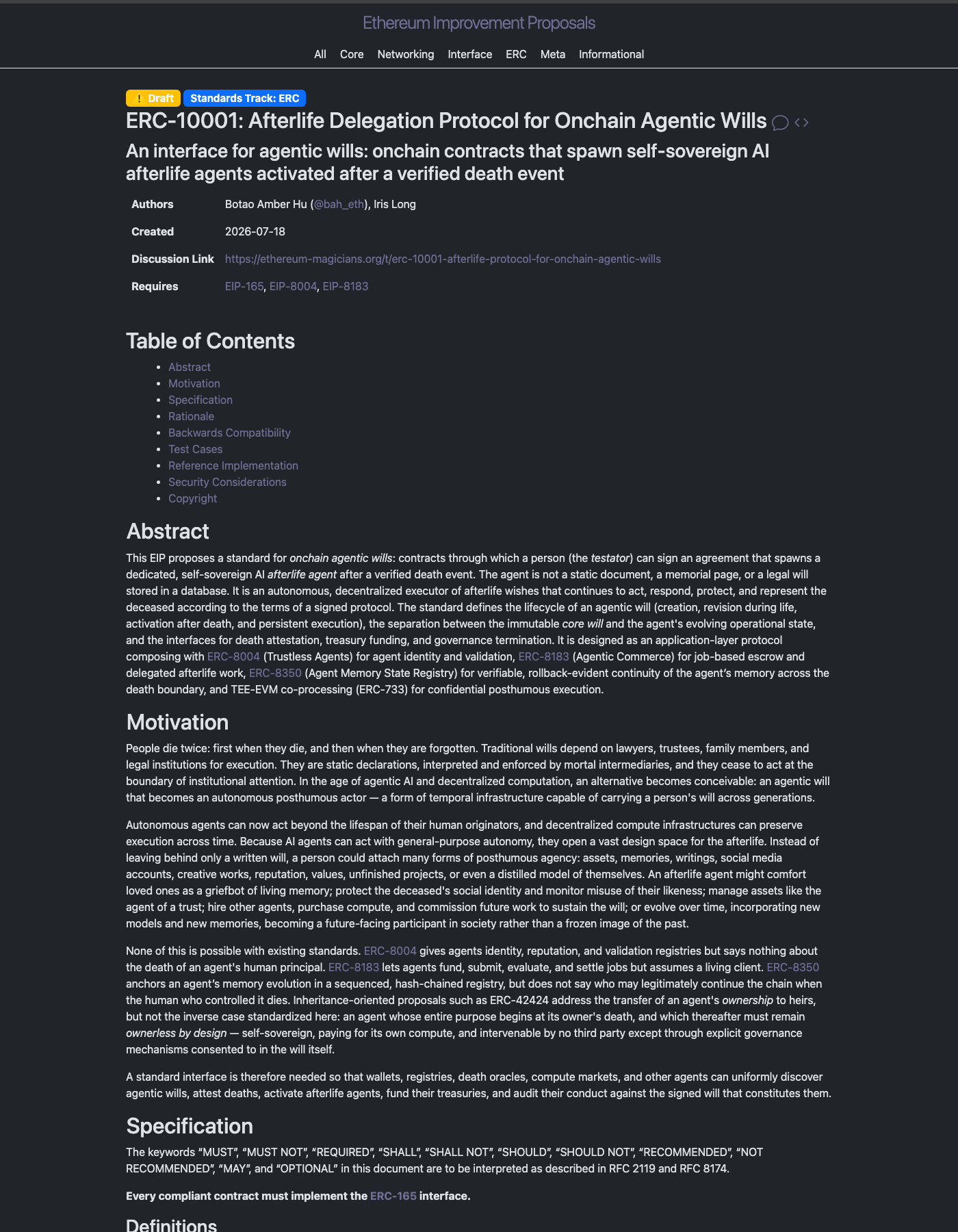}}
\caption{The \texttt{ERC-10001} draft standard as published at \url{https://erc10001.org}, in the
exact format of an Ethereum Improvement Proposal: status badges, front-matter, abstract, and
motivation. The speculative artifact circulates in the genre in which infrastructural futures are
actually negotiated.}
\label{fig:erc10001}
\end{figure}

\subsection{Substrate standards}

\texttt{ERC-10001} is an application-layer protocol composed over four emerging draft standards:

\begin{table}[H]
\centering
\small
\begin{tabular}{p{0.17\linewidth}p{0.36\linewidth}p{0.36\linewidth}}
\toprule
\textbf{Standard} & \textbf{Provides} & \textbf{Role in \texttt{ERC-10001}} \\
\midrule
ERC-733 \citep{erc733tee} & TEE--EVM co-processing: attested confidential computation & The \textit{sealed skin}: the confidential boundary within which the agent's \emph{body} (compute harness) and \emph{mind} (foundation model) run, uninvadable even by its operator, proving on-chain which code ran. \\
ERC-8004 \citep{eip8004} & Trustless-agent identity, reputation, and validation registries & The \textit{identity}: the ID through which the sealed agent exchanges money and information externally, as a discoverable, accountable entity whose behavior third parties audit and score. \\
ERC-8350 \citep{erc8350} & Agent memory state registry: sequenced, hash-chained memory commitments with a strict linearity rule & The \textit{memory}: rollback- and fork-evident memory evolution; controller/authorizer rotation makes death a change of author in one unbroken chain; chain-independent space identity survives chain migration. \\
ERC-8183 \citep{eip8183} & Agentic commerce: job-based escrow (client funds, provider submits, evaluator attests) & The \textit{metabolism}: compute, storage renewal, monitoring, and creative continuation purchased with funds that move only against attested delivery. \\
\bottomrule
\end{tabular}
\caption{The four substrate standards and the organ each supplies.}
\end{table}

\subsection{Lifecycle and constitutional separation}

An agentic will moves through four states --- \texttt{Living} (revisable, revocable),
\texttt{Attested} (death claimed and finalized), \texttt{Activated} (agent enacting), and the
terminal \texttt{Terminated} / \texttt{Revoked}. The core interface comprises
\texttt{createWill}, \texttt{reviseWill}, \texttt{revokeWill}, \texttt{attestDeath},
\texttt{activateAgent}, \texttt{fundTreasury}, and \texttt{terminateAgent}; activation hands the
treasury to the agent rather than to any heir, and termination requires a governance proof
consented to by the principal in the will itself --- a descendant multisig, a court attestation, or
a community vote --- rather than a privileged admin key. The protocol distinguishes the
\textbf{core will} (signed instructions, values, permissions, prohibitions; immutable at death),
\textbf{operational memory} (the ERC-8350 chain of accumulated experience), the \textbf{model
substrate} (upgradeable interpreters constrained by the will), \textbf{delegated subagents}, and
\textbf{commerce records}. The separation prevents an evolving agent from quietly becoming a
different entity: it may adapt its methods, but its legitimacy depends on remaining answerable to
the will that constituted it.

\subsection{Proof of death}

The death oracle is a weighted, contestable proceeding: bonded permissionless claims; evidence
weights and thresholds configured in the will (institutional registries, witness quorums,
dead-man's-switch silence whose weight grows with duration, echoing the legal in-absentia doctrine);
challenge windows scaling inversely with evidence weight; and proof-of-life supremacy --- any
authenticated signal from the principal vetoes all claims and slashes the claimant's bond.
Guardians may suspend proceedings for the incapacitated-not-dead case. The mechanism standardizes
the attestation interface, not the ontology of death; the residual problems (murder incentive
reduced not removed, coercion flagged not prevented, the vanished-but-alive eventually
mis-declared) are stated as explicit non-claims.

\subsection{Enactment loop and sustainability}

The activated agent runs a persistent cycle --- read will, interpret obligations, observe context,
plan, execute, record memory (as ERC-8350 transitions bound to TEE attestations), upgrade or
delegate, sustain itself --- with each epoch committed atomically on-chain: no memory without
attestation, no spending without a memory commit. The treasury sits in a yield-bearing endowment
whose per-epoch outflow is capped at realized yield, with a will-defined austerity ladder (full
operation, memorial-only, archival heartbeat, sealed dormancy) for graceful decline and
permissionless re-funding for revival. If the operator disappears, the agent enters dormancy and
anyone may resurrect it by proving possession of its memories and presenting a fresh attestation,
for an escalating bounty.

\subsection{Open problems}

Deliberately unresolved, as part of the work: proof of death; consent of the living to interactions
authorized by the dead; impersonation (speaking \textit{as} versus \textit{about}); ownership of
shared memories; model drift against an unamendable constitution; asset governance for a posthumous
economic actor; revocation politics; and cultural plurality --- whether a universal protocol can
avoid flattening the traditions through which cultures already tend their dead.

\section{The Afterlife Interviewer}
\label{app:interviewer}

\subsection{Interactive interview}

The Afterlife Interviewer is a language-model agent conducting a long, semi-structured elicitation
in place of a questionnaire. It is built on SparkMe \citep{anugraha2026sparkme}, a multi-agent
adaptive semi-structured interviewing system that conducts multi-turn interviews with strategic
question planning, real-time note-taking, and emergent subtopic discovery. SparkMe decomposes the
interviewer into cooperating roles: a planner that maintains a hierarchical interview plan of main
topics and subtopics and decides, turn by turn, which line of questioning yields the most new
insight; a note-taker that distills each answer into structured notes as the conversation runs; and
the interviewer itself, which phrases the next question conditioned on the plan, the notes, and the
conversational register so far. Crucially, the plan is not fixed: when a participant's answer opens
a theme the protocol's authors did not anticipate, the system promotes it to a subtopic and pursues
it --- the mechanism by which findings like the merit engine or the perpetual penitent
(Section~\ref{sec:results}) surface at all.

We adapt SparkMe by replacing its topic configuration with the will's clause domains --- speech
rights and pronoun policy, funds and their permitted uses, beneficiaries and prohibited
counterparties, memory retention and deletion, termination conditions, drift policy for future
models --- each decomposed into subtopics the planner traverses while adapting to
what the participant has already committed to, pursuing the inconsistencies a fixed instrument
would never surface: the participant who wants their agent silent but also wants it to console
their mother; the participant who forbids it money but assigns it duties that cost money. Because
articulation is iterative, the Interviewer treats contradiction as material rather than error,
returning to unresolved clauses across sessions and staging them in the sandbox rehearsal, where
the participant's own agent confronts scripted situations and the clauses that fail become the next
session's agenda. Sessions run in a web interface with optional voice input and output; per-session
state (plan, notes, transcript) persists per participant so that later sessions resume from the
unresolved clauses rather than restarting the elicitation.

\subsection{Privacy-preserving aggregation}

The corpus concerns grief, private memory, and named third parties; publishing transcripts would be
indefensible. Analysis therefore follows the architecture of Clio \citep{tamkin2024clio}
(\url{https://www.anthropic.com/research/clio}): model-driven extraction of low-dimensional facets
from each interview (clause types, persistence horizons, termination conditions, cosmological
framings), clustering and thematic summarization at the population level, and minimum-aggregation
thresholds so that no theme is reportable unless supported by enough distinct participants to
resist re-identification. Raw transcripts remain sealed; what the research reports is the shape of
the population's answers --- the cross-cultural drift --- never the individual will, which belongs,
as it always has, to its author.

\end{document}